 \documentclass[10pt,pra,aps,a4paper,oneside,twocolumn,superscriptaddress,showpacs]{revtex4-1}

\usepackage[english]{babel}
\usepackage[latin1]{inputenc}
\usepackage[T1]{fontenc}
\usepackage{lmodern}
\usepackage{ulem}
\usepackage{dcolumn}
\usepackage{subfigure}
\usepackage{footnote}
\usepackage{multirow}
\usepackage{amsmath,amsfonts,amssymb}

\usepackage{graphicx}  
\usepackage{latexsym}   
\usepackage{color}

\begin{document}

\title{Electromagnetic energy within coated spheres containing dispersive metamaterials}

\author{\firstname{Tiago}  J.  \surname{Arruda}}
\email{tiagojarruda@gmail.com}
\affiliation{Faculdade de Filosofia,~Ci\^encias e Letras de Ribeir\~ao Preto, Universidade de S\~ao Paulo,\\
 14040-901 Ribeir\~ao Preto, S\~ao Paulo, Brazil}

\author{\firstname{Felipe}  A. \surname{Pinheiro}}
\affiliation{Instituto de F\'{i}sica, Universidade Federal do Rio de Janeiro, 21941-972 Rio de Janeiro-RJ, Brazil}

\author{\firstname{Alexandre} S. \surname{Martinez}}
\affiliation{Faculdade de Filosofia,~Ci\^encias e Letras de Ribeir\~ao Preto, Universidade de S\~ao Paulo,\\
 14040-901 Ribeir\~ao Preto, S\~ao Paulo, Brazil}
\affiliation{National Institute of Science and Technology in
Complex Systems, 22290-180 Rio de Janeiro-RJ, Brazil}

\begin{abstract}

An exact expression is derived for the time-averaged electromagnetic energy within a magneto-dielectric coated sphere, which is irradiated by a plane and time-harmonic electromagnetic wave.
Both the spherical shell and core are considered to be dispersive and lossy, with a realistic dispersion relation of an isotropic split-ring resonator metamaterial.
We obtain analytical expressions for the stored electromagnetic energies inside the core and the shell separately and calculate their contribution to the total average energy density.
The stored electromagnetic energy is calculated for two situations involving a metamaterial coated sphere: the dielectric shell and dispersive metamaterial core, and vice-versa.
An explicit relation between the stored energy and the optical absorption efficiency is also obtained.
We show that the stored electromagnetic energy is an observable sensitive to field interferences responsible for the Fano effect. This result, together with the fact that the Fano effect is more likely to occur in metamaterials with negative refraction, suggest that our findings may be explored in applications.

\end{abstract}


\maketitle


\section{Introduction}

Electromagnetic (EM) scattering by small particles is a fundamental topic in classical electrodynamics.
It has a broad range of applications, including biology, meteorology, astronomy, and medicine.
Historically, a complete solution for homogeneous spheres with arbitrary size was first derived, in an independent way, by Lorenz and Mie more than a century ago~\cite{Bohren,Hulst}.
For this reason, this solution is today widely known as Lorenz-Mie solution.
Despite its long history, the research on EM scattering still reveals surprises.
Giant resonances that anomalously increase with the resonances order (dipole, quadrupole, etc.)~\cite{Tribelsky}, the formation of complex field structures with vortices inside scattering particles~\cite{Bashevoy}, and the superscattering of light in subwavelength structures, in which the single-channel limit can be overcome~\cite{Ruan}, are some examples of interesting, unexpected, and basic phenomena that have been recently unveiled in the field of EM scattering.

Most of these recent results have been driven by the extraordinary technological progresses in the emerging field of nanophotonics.
In this field, the understanding of the interaction between EM radiation and individual nanostructures is crucial.
Indeed, one of the most important challenges is to confine light at the subwavelength scale, leading to an enhancement of the EM field.
The excitation of plasmons localized on the surface of nanoparticles is a common strategy to achieve such enhancement~\cite{Maier}.
The investigation of EM scattering by metallic nanoparticles provides important insights on the understanding and control of the excitation of surface plasmons. It allows the optimization of the field enhancement, and hence the design of novel photonic devices.
One important example is the metallic coated spherical nanoparticle.
The Lorenz-Mie-type solution for this geometry, first obtained by Aden and Kerker~\cite{Aden}, provides the theoretical basis for many modern applications, such as the off-resonance field enhancement in EM scattering~\cite{Miroshnichenko2010} and the spaser-based nanolaser~\cite{Spaser}, for instance.

In the last decade, the development of metamaterials has opened up new frontiers in photonics and plasmonics.
Metamaterials exhibit unusual optical properties, with no counterpart in natural media, that can be exploited to generate negative refraction~\cite{Smith00}, resolve images beyond the diffraction limit~\cite{Smith2004}, exhibit optical magnetism~\cite{Enkrich2005,Cai2007}, and act as an EM cloak~\cite{Pendry2006,Leonhardt2006}.
As many fundamental aspects of classical electrodynamics, EM scattering has been revised due the advent of metamaterials and the possibility of negative refraction~\cite{Ruppin-negative,Gao,Liu2004}.
In particular, the problem of EM scattering by spheres with negative refraction has been addressed in~\cite{Liu2004}.
The related problem of the EM energy density stored in dispersive metamaterials, including unavoidable losses and realistic effective constitutive parameters, has been recently treated for coated cylinders~\cite{Tretyakov2005}, split-ring resonators (SRR)~\cite{Tretyakov,Boardman,Luan1}, and chiral metamaterials~\cite{Luan2}.
As far as we are aware, the case of the EM energy density within coated spheres, an important geometry for applications in plasmonics, made of metamaterials has never been treated so far.

The aim of the this paper is to fill this gap by investigating the EM energy density within coated spheres made of dispersive metamaterials, including losses and realistic effective parameters for wires-SRR metamaterials.
To accomplish this, we use the magnetic Aden-Kerker solution~\cite{Bohren} for the internal EM field to obtain an exact expression for the time-averaged EM energy within a dispersive spherical shell and core.
This generalizes the results already obtained for homogeneous spheres irradiated by plane waves~\cite{Bott,Ruppin-sphere,Tiago-sphere,Ruppin-energy}, and allows us to analyze the resonances inside the core and shell separately.

This paper is organized as follows.
In Sec.~\ref{mie}, the expressions of the internal EM field inside the magnetic coated sphere are presented.
Section~\ref{time} is devoted to the calculation of the time-averaged EM energy density, the main result of this paper.
The dispersion relations associated with a wires-SRR metamaterial are presented in Sec.~\ref{dispersive}.
Section~\ref{numerical} is reserved to the numerical results whereas in Sec.~\ref{conclusions} we provide a summary of our main conclusions.

\section{Aden-Kerker solution}
\label{mie}

Let $(\mathbf{E},\mathbf{H})$ be a plane and complex EM wave with time-harmonic dependence $\exp({-{\rm i} \omega t})$, where $\omega$ is the angular frequency, and electric field amplitude is $E_0$.
This wave is incident on a magneto-dielectric coated sphere with inner radius $a$ and outer one $b$.
The involved media are assumed to be linear, homogeneous and isotropic, with scalar electric permittivity and magnetic permeability $(\epsilon_1,\mu_1)$ for the core ($0\leq r\leq a$), $(\epsilon_2,\mu_2)$ for the spherical shell ($a\leq r\leq b$), and $(\epsilon_0,\mu_0)$ for the surrounding medium ($r\geq b$), with the last one considered to be the free-space~\cite{Bohren}.
Under these assumptions, the macroscopic Maxwell's equations for non-optically active media provide the vector Helmholtz equation
$[\mathbf{\nabla}^2+k^2](\mathbf{E},\mathbf{H})=(\mathbf{0},\mathbf{0})$, where $k=2\pi/\lambda$ is the wave number and $\lambda$ is the wavelength of the radiation in the respective medium.
The solutions of this equation for each region above delimited, which are a generalization of the Lorenz-Mie solutions~\cite{Bohren,Hulst}, are known as the Aden-Kerker solutions~\cite{Bohren,Aden}.
Solving this vector equation, the components of the internal EM fields ($\mathbf{E}_q,\mathbf{H}_q$) in the spherical coordinate system $(r,\theta,\phi)$ inside the spherical core ($q=1$) and shell ($q=2$) are explicitly given~\cite{Bohren,Aden,Tiago-sphere}:
\begin{eqnarray}
        E_{qr}&=&-\displaystyle \frac{{\rm i} \cos\phi\sin\theta}{\rho_q^2}\sum_{n=1}^{\infty}E_nn(n+1)\pi_n\{\delta_{1,q}d_n\psi_n(\rho_q)\nonumber\\
        &&\qquad\qquad\quad +\delta_{2,q}\left[g_n\psi_n(\rho_q)-w_n \chi_n(\rho_q)\right]\}\ ,\label{Eq-r}\\
        E_{q\theta}&=&\displaystyle\frac{\cos\phi}{\rho_q}\sum_{n=1}^{\infty}E_n\big(\delta_{1,q}\left[c_n \pi_n\psi_n(\rho_q)-{\rm i} d_n \tau_n\psi_n'(\rho_q)\right]\nonumber\\
        &&\qquad\qquad\quad +\delta_{2,q}\{\pi_n\left[f_n \psi_n(\rho_q)-v_n \chi_n(\rho_q)\right]\nonumber\\
        &&\qquad\qquad\qquad -{\rm i}\tau_n\left[g_n \psi_n'(\rho_q)-w_n \chi_n'(\rho_q)\right]\}\big)\ ,\\
        E_{q\phi}&=&\displaystyle\frac{\sin\phi}{\rho_q}\sum_{n=1}^{\infty}E_n\big(\delta_{1,q}\left[{\rm i} d_n \pi_n\psi_n'(\rho_q)-c_n \tau_n\psi_n(\rho_q)\right]\nonumber\\
        &&\qquad\qquad\quad+ \delta_{2,q}\{{\rm i}\pi_n\left[g_n \psi_n'(\rho_q)-w_n \chi_n'(\rho_q)\right]\nonumber\\
        &&\qquad\qquad\qquad -\tau_n\left[f_n \psi_n(\rho_q)-v_n \chi_n(\rho_q)\right]\} \big)\ ,\label{Eq-phi}
        \end{eqnarray}
        \begin{eqnarray}
        H_{qr}&=&-\displaystyle\frac{ {\rm i} k_q\sin\phi\sin\theta}{\omega\mu_q\rho_q^2}\sum_{n=1}^{\infty}E_nn(n+1)\pi_n\{\delta_{1,q} c_n\psi_n(\rho_q)\nonumber\\
        &&\qquad\qquad\quad +\delta_{2,q}\left[f_n\psi_n(\rho_q)-v_n\chi_n(\rho_q)\right]\}\ ,\label{Hq-r}\\
        H_{q\theta}&=&\displaystyle\frac{k_q\sin\phi}{\omega\mu_q\rho_q}\sum_{n=1}^{\infty}E_n\big(\delta_{1,q}\left[d_n \pi_n\psi_n(\rho_q)-{\rm i} c_n \tau_n\psi_n'(\rho_q)\right]\nonumber\\
        &&\qquad\qquad\quad+\delta_{2,q}\{\pi_n\left[g_n \psi_n(\rho_q)-w_n \chi_n(\rho_q)\right]\nonumber\\
        &&\qquad\qquad\qquad -{\rm i}\tau_n\left[f_n \psi_n'(\rho_q)-v_n
        \chi_n'(\rho_q)\right]\}\big)\ ,\\
        H_{q\phi}&=&\displaystyle\frac{k_q\cos\phi}{\omega\mu_q\rho_q}\sum_{n=1}^{\infty}E_n\big(\delta_{1,q}\left[d_n \tau_n\psi_n(\rho_q)-{\rm i} c_n \pi_n\psi_n'(\rho_q)\right]\nonumber\\
        &&\qquad\qquad\quad+\delta_{2,q}\{\tau_n\left[g_n \psi_n(\rho_q)-w_n \chi_n(\rho_q)\right]\nonumber\\
        &&\qquad\qquad\qquad-{\rm i}\pi_n\left[f_n \psi_n'(\rho_q)-v_n \chi_n'(\rho_q)\right]\}\big)\ ,\label{Hq-phi}
\end{eqnarray}
where $\rho_q=k_qr$, with $k_q$ being the wave number in the medium $(\epsilon_q,\mu_q)$, $\delta_{q,q'}$ is the Kronecker delta and $E_n={\rm i}^nE_0({2n+1})/[{n(n+1)}]$~\cite{Bohren}.
The radial functions $\psi_n(\rho_q)=\rho_q j_n(\rho_q)$ and $\chi_n(\rho_q)=-\rho_q y_n(\rho_q)$ are the Riccati-Bessel and Riccati-Neumann functions, respectively, where $j_n$ is the spherical Bessel function and $y_n$ is the Neumann function.
The angular functions are $\pi_n={P_n^1(\cos\theta)}/{\sin\theta}$ and $\tau_n={{\rm d}}[{P_n^1(\cos\theta)}]/{{\rm d}\theta}$, where $P_n^1$ is the associated Legendre function of first order.
The expressions for the incident and scattered EM fields in terms of vector spherical harmonics and the scattering coefficients $a_n$ and $b_n$ can be found in~\cite{Bohren}.

Imposing the continuity of the tangential components of the EM fields at the interfaces ($r=a$ and $r=b$) between the media, we obtain the Aden-Kerker coefficients in the magnetic case ($\mu_q\not=\mu_0$)~\cite{Bohren,Aden}:
    \begin{eqnarray}
        a_n &=&\frac{\widetilde{m}_2\psi_n'(y)\alpha_n-\psi_n(y)\widetilde{\alpha}_n}{\widetilde{m}_2\xi_n'(y)\alpha_n-\xi_n(y)\widetilde{\alpha}_n}\
        ,\label{an}\\
        b_n &=&\frac{\psi_n'(y)\beta_n-\widetilde{m}_2\psi_n(y)\widetilde{\beta}_n}{\xi_n'(y)\beta_n-\widetilde{m}_2\xi_n(y)\widetilde{\beta}_n}\
        ,\label{bn}\\
        c_n &=&\frac{m_1f_n\left[\psi_n(m_2x)-B_n\chi_n(m_2x)\right]}{m_2\psi_n(m_1x)}\
        ,\\
        d_n &=&\frac{m_1g_n\left[\psi_n'(m_2x)-A_n\chi_n'(m_2x)\right]}{m_2\psi_n'(m_1x)}\
        ,\\
        f_n &=&\frac{m_2{\rm i}}{\xi_n'(y)\beta_n-\widetilde{m}_2\xi_n(y)\widetilde{\beta}_n}\
        ,\\
        g_n&=&\frac{m_2{\rm i}}{\widetilde{m}_2\xi_n'(y)\alpha_n-\xi_n(y)\widetilde{\alpha}_n}\
        ,\\
        v_n &=&B_nf_n \ ,\\
        w_n &=&A_ng_n \ ,
    \end{eqnarray}
with the auxiliary functions expressed by
    \begin{eqnarray*}
        A_n &=& \frac{\widetilde{m}_2\psi_n(m_2x)\psi_n'(m_1x)-\widetilde{m}_1\psi_n'(m_2x)\psi_n(m_1x)}{\widetilde{m}_2\chi_n(m_2x)\psi_n'(m_1x)-\widetilde{m}_1\chi_n'(m_2x)\psi_n(m_1x)}\
        ,\\
        B_n&=&\frac{\widetilde{m}_2\psi_n'(m_2x)\psi_n(m_1x)-\widetilde{m}_1\psi_n(m_2x)\psi_n'(m_1x)}{\widetilde{m}_2\chi_n'(m_2x)\psi_n(m_1x)-\widetilde{m}_1\chi_n(m_2x)\psi_n'(m_1x)}\
        ,\\
        \alpha_n &=& \psi_n(m_2y)-A_n\chi_n(m_2y)\ ,\\
        \beta_n &=& \psi_n(m_2y)-B_n\chi_n(m_2y)\ , \\
        \widetilde{\alpha}_n &=& \psi_n'(m_2y)-A_n\chi_n'(m_2y)\ ,\\
        \widetilde{\beta}_n &=& \psi_n'(m_2y)-B_n\chi_n'(m_2y)\ .
    \end{eqnarray*}
The quantities $x=ka$ and $y=kb$ are the size parameters related to the inner and the outer spheres, respectively, with $k$ being the incident wave number.
The function $\xi_n(\rho_q)=\psi_n(\rho_q)-{\rm i}\chi_n(\rho_q)$ is the Riccati-Hankel function.
The refractive and impedance indices are $m_q=k_q/k=[{\epsilon_q\mu_q/(\epsilon_0\mu_0)}]^{1/2}$ and $\widetilde{m}_q=\mu_0 m_q/\mu_q=[{\epsilon_q\mu_0/(\epsilon_0\mu_q)}]^{1/2}$~\cite{Kerker:1983,Alexandre-new,Alexandre-vanish}, respectively (with $q=1$ for the core and $q=2$ for the shell).

\section{Time-averaged internal energy}
\label{time}

Consider a time-harmonic EM field $(\mathbf{E}_{q},\mathbf{H}_{q})$ confined to a homogeneous and isotropic medium which is restricted to a spherical shell ($l_1\leq r\leq l_2$), with complex optical constants $\epsilon_{q}=\epsilon_q'+{\rm i}\epsilon_q''$ and $\mu_q=\mu_q'+{\rm i}\mu_q''$.
The time-averaged EM energy ($W_{q}=\langle W_q\rangle_t$) stored inside this region can be calculated by means of the equation \cite{Landau}
\begin{eqnarray}
    W_{q}(l_1,l_2)=\int_0^{2\pi} {\rm d}\phi \int_{-1}^{1} {\rm
    d}(\cos\theta) \int_{l_1}^{l_2} {\rm d}r\ r^2\langle u_q\rangle_t\ ,\label{energy}
\end{eqnarray}
where $\langle u_q\rangle_t=\langle u_q\rangle_t(r,\cos\theta,\phi)$ is the corresponding time-averaged energy density:
\begin{eqnarray}
    \langle u_q\rangle_t=\frac{1}{4}\left[{\epsilon_{q}^{\rm(eff)}}\left|\mathbf{E}_{q}\right|^2+{\mu_{q}^{\rm(eff)}}\left|\mathbf{H}_{q}\right|^2\right]\ ,\label{density}
\end{eqnarray}
with $\epsilon_q^{\rm(eff)}$ and $\mu_q^{\rm(eff)}$ being the effective electric and magnetic energy coefficients, respectively~\cite{Boardman}.
Specially, for dispersive and weakly absorbing medium ($\epsilon_q'\gg\epsilon_q''$, $\mu_q'\gg\mu_q''$), one has the positive definite coefficients~\cite{Landau}
\begin{eqnarray}
    \epsilon_q^{\rm(eff)}(\omega)&=&\frac{\partial\left[\omega\epsilon_q'(\omega)\right]}{\partial\omega}>0\
    ,\label{eps-lan}\\
    \mu_q^{\rm(eff)}(\omega)&=&\frac{\partial\left[\omega\mu_q'(\omega)\right]}{\partial\omega}>0\
    . \label{mu-lan}
\end{eqnarray}
If the lossless medium $(\epsilon_q,\mu_q)$ is non-dispersive, one readily obtains $\epsilon_q^{\rm(eff)}=\epsilon_q'$ and $\mu_q^{\rm(eff)}=\mu_q'$ from equations~(\ref{eps-lan}) and (\ref{mu-lan})~\cite{Bott,Landau}.
To guarantee the positiveness of the EM energy density, this last result imposes that materials with both $\epsilon_q'<0$ and $\mu_q'<0$ are necessarily dispersive.
We stress that, for lossy and dispersive materials, however, the quantities $[\epsilon_q^{\rm(eff)},\mu_q^{\rm(eff)}]$ in equation~(\ref{density}) cannot be calculated using equations~(\ref{eps-lan}) and (\ref{mu-lan}).
Indeed, these quantities depend on the approach and the model used to describe the dispersion relations~\cite{Boardman,Landau}.

If the shell has the same optical properties as the surrounding medium $(\epsilon_0,\mu_0)$, which is assumed to be non-dispersive and non-absorbing, one has
\begin{eqnarray}
    W_{0}(l_1,l_2) = \frac{2}{3}\pi|E_0|^2\epsilon_0(l_2^3-l_1^3)\
    .\label{w0}
\end{eqnarray}

In addition, for $m_{q}\not=m_{q}^*$ ($q=\{1,2\}$), one has an analytical expression for the integral that appears in the radial part of equation~(\ref{energy}), which involves product of the spherical Bessel and Neumann functions~\cite{Watson}.
For sake of simplicity, we define the dimensionless function~\cite{Tiago-sphere,Ruppin-energy,Tiago-cylinder}
\begin{eqnarray}
    &&\frac{\mathcal{I}_{q,n}^{(z\bar{z})}(l_1,l_2)}{1/(l_2^3-l_1^3)}=\int_{l_1}^{l_2}{\rm
    d}r\ r^2z_n(\rho_{q})\bar{z}_n(\rho_{q}^*)\nonumber\\
    &=&r^3\frac{\left[\rho_{q}^*z_n(\rho_{q})\bar{z}_n'(\rho_{q}^*)-\rho_{q}z_n'(\rho_{q})\bar{z}_n(\rho_{q}^*)\right]}{\rho_{q}^2-\rho_{q}^{*2}}\Bigg|_{r=l_1}^{r=l_2}\
    ,\label{integral1}
\end{eqnarray}
where $z_n$ e $\bar{z}_n$ may be any spherical Bessel or Neumann functions, and $l_1,l_2\in\mathbb{R}$ are the limits of integration.
In particular, if $m_{q}=\pm m_{q}^*$ ($q=\{1,2\}$), using L'Hospital's rule and recurrence relations involving spherical Bessel and Neumann functions~\cite{Watson}, we obtain
\begin{eqnarray}
    \frac{\mathcal{I}_{q,n}^{(z\bar{z})}(l_1,l_2)}{1/(l_2^3-l_1^3)}&=&\lim_{m_q\to\pm m_q^*}\int_{l_1}^{l_2}{\rm
    d}r\ r^2z_n(m_qkr)\bar{z}_n(m_q^*kr)\nonumber\\
    &=&\varrho_{\pm,n}^{(z\bar{z})}\frac{r^3}{4}\big[2z_n(\rho_{q})\bar{z}_n(\rho_{q})-z_{n-1}(\rho_{q})\bar{z}_{n+1}(\rho_{q})\nonumber\\
    && \qquad-z_{n+1}(\rho_{q})\bar{z}_{n-1}(\rho_{q})\big]\bigg|_{r=l_1}^{r=l_2}\
    ,\label{integral2}
\end{eqnarray}
where one must necessarily choose $\varrho_{+,n}^{(z\bar{z})}=1$, for $m_q=m_q^*$ [{i.e.}, ${\rm Im}(m_q)=0$].
For $m_q=-m_q^*$ [${\rm Re}(m_q)=0$], according to the definition in equation~(\ref{integral2}), one has the following combinations: $\varrho_{-,n}^{(jj)}=\varrho_{-,n}^{(yj)}=(-1)^n$ and $\varrho_{-,n}^{(yy)}=\varrho_{-,n}^{(jy)}=(-1)^{n+1}$, since $j_n(-\rho)=(-1)^nj_n(\rho)$ and $y_n(-\rho)=(-1)^{n+1}y_n(\rho)$~\cite{Watson}.
Equations~(\ref{integral1}) and (\ref{integral2}) are quite suitable to simplify the expressions associated with the average internal energy~\cite{Bott,Tiago-sphere,Ruppin-energy,Tiago-cylinder} and intensities~\cite{Kaiser}, and it is the first time that these general formulations, including both Bessel and Neumann functions, are used in this context of magnetic scatterers.

In the spherical core $(0\leq r\leq a)$, separating the electric and magnetic fields contributions to the average EM energy, that is, $W_{1E}\equiv\int_{\mathcal{V}_1}{\rm d}^3r \epsilon_1^{\rm(eff)}|\mathbf{E}_{1}|^2/4$ and $W_{1H}\equiv\int_{\mathcal{V}_1}{\rm d}^3r \mu_1^{\rm(eff)}|\mathbf{H}_{1}|^2/4$, with $\mathcal{V}_1$ being the respective region of integration, we obtain, respectively:
\begin{widetext}
\begin{eqnarray}
    \frac{W_{1E}(0,a)}{W_0(0,a)}&=&\frac{3}{4}\frac{\epsilon_1^{\rm(eff)}}{\epsilon_0}\sum_{n=1}^{\infty}\bigg\{(2n+1) |c_n |^2 \mathcal{I}_{1,n}^{(jj)}(0,a)+ |d_n |^2\left[n \mathcal{I}_{1,n+1}^{(jj)}(0,a)+(n+1)\mathcal{I}_{1,n-1}^{(jj)}(0,a)
    \right]\bigg\}\ ,\label{w1-e}\\
    \frac{W_{1H}(0,a)}{W_0(0,a)}&=&\frac{3}{4}\left|\widetilde{m}_1\right|^2\frac{\mu_1^{\rm(eff)}}{\mu_0}\sum_{n=1}^{\infty}\bigg\{(2n+1) |d_n |^2 \mathcal{I}_{1,n}^{(jj)}(0,a)+|c_n |^2\left[n\mathcal{I}_{1,n+1}^{(jj)}(0,a)+(n+1)\mathcal{I}_{1,n-1}^{(jj)}(0,a)
    \right]\bigg\}\ ,\label{w1-h}
\end{eqnarray}
\end{widetext}
where we have used equations~(\ref{Eq-r})--(\ref{Hq-phi}), (\ref{energy}), (\ref{w0}) and (\ref{integral1}) [or (\ref{integral2})] for $q=1$, $l_1=0$ and $l_2=a$.
The radial and angular contributions have been simplified by applying the relations $({2n+1})\int_{-1}^{1} {\rm d}(\cos \theta) \pi_n   \pi_{n'}\sin^2 \theta  = {2n(n+1)}\delta_{n,n'}$, $({2n+1})\int_{-1}^{1} {\rm d}(\cos \theta) (\pi_n   \pi_{n'} + \tau_n  \tau_{n'} ) =  {2n^2(n+1)^2}\delta_{n,n'}$
and $\int_{-1}^{1} {\rm d}(\cos \theta) (\pi_n \tau_{n'} + \tau_n    \pi_{n'} ) = 0$~\cite{Bohren,Tiago-sphere}.
Therefore, from equations~(\ref{w1-e}) and (\ref{w1-h}), we calculate the average EM energy within the core as the sum of the electric and magnetic contributions:
\begin{eqnarray}
    W_1(0,a)=W_{1E}(0,a)+W_{1H}(0,a)\ .\label{W1}
\end{eqnarray}
These results are in agreement with the papers of Bott and Zdunkowski~\cite{Bott} in the nonmagnetic case ($\mu_1=\mu_0$) and Ruppin~\cite{Ruppin-energy} for a Lorentz-type permeability.
Some details of these calculations for the magnetic case ($\mu_1\not=\mu_0$) can be found in~\cite{Tiago-sphere}.

Analogously, in the shell region ($a\leq r\leq b$) we obtain the electric and magnetic contributions to the average EM energy, respectively:
\begin{widetext}
\begin{eqnarray}
    \frac{W_{2E}(a,b)}{W_0(a,b)}&=&\frac{3}{4}\frac{\epsilon_2^{\rm(eff)}}{\epsilon_0}\sum_{n=1}^{\infty}\Bigg\{(2n+1) |f_n |^2 \mathcal{I}_{2,n}^{(jj)}(a,b)+ |g_n |^2 \left[n \mathcal{I}_{2,n+1}^{(jj)}(a,b)+(n+1)\mathcal{I}_{2,n-1}^{(jj)}(a,b)\right]\nonumber\\
    &&+ (2n+1) |v_n |^2 \mathcal{I}_{2,n}^{(yy)}(a,b)+ |w_n |^2 \left[n \mathcal{I}_{2,n+1}^{(yy)}(a,b)+(n+1)\mathcal{I}_{2,n-1}^{(yy)}(a,b)\right]\nonumber\\
    &&+ 2{\rm Re}\bigg[(2n+1) f_n v_n^{*} \mathcal{I}_{2,n}^{(jy)}(a,b)+ g_n w_n^{*}\left(n \mathcal{I}_{2,n+1}^{(jy)}(a,b)+(n+1)\mathcal{I}_{2,n-1}^{(jy)}(a,b)\right)\bigg]\Bigg\}\ ,\label{w2-e}
\end{eqnarray}
\begin{eqnarray}
    \frac{W_{2H}(a,b)}{W_0(a,b)}&=&\frac{3}{4}\left|\widetilde{m}_2\right|^2\frac{\mu_2^{\rm(eff)}}{\mu_0}\sum_{n=1}^{\infty}\Bigg\{(2n+1) |g_n |^2 \mathcal{I}_{2,n}^{(jj)}(a,b)+ |f_n |^2 \left[n \mathcal{I}_{2,n+1}^{(jj)}(a,b)+(n+1)\mathcal{I}_{2,n-1}^{(jj)}(a,b)\right]\nonumber\\
    && + (2n+1) |w_n |^2 \mathcal{I}_{2,n}^{(yy)}(a,b)+ |v_n |^2 \left[n \mathcal{I}_{2,n+1}^{(yy)}(a,b)+(n+1)\mathcal{I}_{2,n-1}^{(yy)}(a,b)\right]\nonumber\\
    && + 2{\rm Re}\bigg[(2n+1) g_n w_n^{*} \mathcal{I}_{2,n}^{(jy)}(a,b)+ f_n v_n^{*}\left(n \mathcal{I}_{2,n+1}^{(jy)}(a,b)+(n+1)\mathcal{I}_{2,n-1}^{(jy)}(a,b)\right)\bigg]\Bigg\}\ ,\label{w2-h}
\end{eqnarray}
\end{widetext}
where we have employed equations~(\ref{Eq-r})--(\ref{Hq-phi}), (\ref{energy}), (\ref{w0}) and (\ref{integral1}) [or (\ref{integral2})] for $q=2$, $l_1=a$ and $l_2=b$.
The average EM energy within the spherical shell is readily calculated by the expression
\begin{eqnarray}
    W_2(a,b)=W_{2E}(a,b)+W_{2H}(a,b)\ .\label{W2}
\end{eqnarray}

From equations~(\ref{W1}) and (\ref{W2}), we finally obtain the time-averaged EM energy within the magneto-dielectric coated sphere as the sum of the average energies stored in the spherical core and shell:
\begin{eqnarray}
    W_{1;2}(a,b)=W_1(0,a)+W_2(a,b)\ .   \label{W12}
\end{eqnarray}
Therefore, given the constitutive parameters associated with the media $(\epsilon_q,\mu_q)$, $q=\{1,2\}$, one can readily determine the respective average EM energy inside the spherical particle by means of equations~(\ref{w1-e})--(\ref{W12}).

\section{Dispersive metamaterial}
\label{dispersive}

A composite isotropic metamaterial $(\epsilon_q,\mu_q)$, consisting of an array of wires and an array of split-ring resonators (SRR), can be described by the {effective} scalar quantities~\cite{Smith00}
\begin{eqnarray}
    {\epsilon_q}(\omega)&=&{\epsilon_0}\left[1-\frac{\omega_{\rm p}^2}{\omega(\omega+{\rm i}\gamma)}\right]\;,\label{eps1}\\
    {\mu_q}(\omega)&=&{\mu_0}\left[1-\frac{F\omega^2}{\left(\omega^2-\omega_0^2\right)+{\rm i}\omega\Gamma}\right]\;,\label{mu1}
\end{eqnarray}
where $\omega_{\rm p}$ and $\omega_0$ are the effective plasma and resonance frequencies associated with the wire and the SRR media, respectively.
The dimensionless factor $F$ is the fractional area of the unit cell occupied by the interior of the split-ring, and $\gamma$ and $\Gamma$ are damping coefficients.
For a certain frequency band, at the microwave range, the real part of the dispersive quantities in equations~(\ref{eps1}) and (\ref{mu1}) are both negative $(\epsilon_q'<0,\mu_q'<0)$.
These negative parameters lead to a negative index of refraction and the metamaterial is considered to be ``left-handed" because, for plane waves propagating through this medium at this frequency range, the wavevector lies in the direction opposite to the EM energy flux, given by the Poynting vector~\cite{Soukoulisbook}.
In the lossless situation, one has explicitly the refractive index $m_q=({\epsilon_q/\epsilon_0})^{1/2}({\mu_q/\mu_0})^{1/2}=p[{\epsilon_q\mu_q/(\epsilon_0\mu_0)}]^{1/2}$, where $p=-1$ if both $\epsilon_q'$ and $\mu_q'$ are negative, and $p=1$ otherwise.
It is important to emphasize that metamaterials exhibiting negative refraction must be dispersive to guarantee the positiveness of the EM energy density~\cite{Soukoulisbook}.

From the non-Lorentz-type model provided by the SRR media~\cite{Smith00}, equation~(\ref{mu1}), and the plasma-like dispersion associated with the wires, equation~(\ref{eps1}), Tretyakov~\cite{Tretyakov} has determined an exact expression for the EM energy density using an equivalent circuit (EC) approach.
Both the EC and the electrodynamic (ED) approaches have been discussed by Boardman and Marinov~\cite{Boardman} who, inspired by an early work by Loudon~\cite{Loudon}, consider the limits of validity of the energy density in metamaterials of Lorentz- and SRR-type.
The effective electric and magnetic energy coefficients, which enter equation~(\ref{density}), are, in the ED approach, given by~\cite{Boardman}:
${\epsilon_q^{\rm( eff)}}|_{\rm ED}=\epsilon_0[1+{\omega_{\rm p}^2}/({\omega^2+\gamma^2})]$ and ${\mu_q^{\rm(eff)}}|_{\rm ED}=\mu_0(1+{F\omega^2[\omega_0^2(3\omega_0^2-\omega^2)+\omega^2\Gamma^2]}/\{{\omega_0^2[(\omega_0^2-\omega^2)^2+\omega^2\Gamma^2]}\})$.
As pointed out in~\cite{Boardman}, these energy coefficients in ED approach are in perfect agreement in dispersive and lossless materials ($\gamma=\Gamma=0$), described in~\cite{Landau} by equations~(\ref{eps-lan}) and (\ref{mu-lan}).
In EC approach, the effective energy coefficients calculated by Tretyakov~\cite{Tretyakov} are $\epsilon_q^{\rm(eff)}|_{\rm EC}=\epsilon_q^{\rm(eff)}|_{\rm ED}$ and
${\mu_q^{\rm(eff)}}|_{\rm EC}=\mu_0\{1+{F\omega^2(\omega_0^2+\omega^2)}/[{(\omega_0^2-\omega^2)^2+\omega^2\Gamma^2}]\}$,
which are more adequate to describe the low-frequency range $\omega<\omega_0$ than the magnetic energy coefficient provided by the Lorentz-type model~\cite{Ruppin-energy,Ruppin-dispersive}: ${\mu_q}|_{\rm Lorentz}=\mu_0\{1-{F\omega_0^2}/[{(\omega^2-\omega_0^2)+{\rm i}\omega\Gamma}]\}$ and $\mu_q^{\rm(eff)}|_{\rm Lorentz}=\mu_0\{1+{F\omega_0^2(\omega^2+\omega_0^2)}/[{(\omega_0^2-\omega^2)^2+\omega^2\Gamma^2}]\}$.
On the other hand, the Lorentz-type dispersion is a more adequate description than the EC approach in the high-frequency range $\omega>\omega_0$~\cite{Boardman}.
Both approaches, however, do not satisfy the classical energy formula for dispersive lossless materials in time-harmonic fields.

Recently, Luan~\cite{Luan1} has shown that if the power loss is a priori identified, the EC approach reduces to the ED one, and the correct expressions for the effective electric and magnetic energy coefficients, which are consistent with equations~(\ref{eps-lan}) and (\ref{mu-lan}), are
\begin{eqnarray}
    {\epsilon_q^{\rm( eff)}}(\omega)&=&{\epsilon_0}\left[1+\frac{\omega_{\rm
    p}^2}{\omega^2+\gamma^2}\right]\ ,\label{eps-eff}\\
    {\mu_q^{\rm(eff)}}(\omega)&=&{\mu_0}\left[1+\frac{F\omega^2\left(3\omega_0^2-\omega^2\right)}{\left(\omega_0^2-\omega^2\right)^2+\omega^2\Gamma^2}\right]\
.\label{mu-eff}
\end{eqnarray}

In the following, we employ the constitutive quantities calculated by Luan~\cite{Luan1}, equations~(\ref{eps-eff}) and (\ref{mu-eff}), since they are valid for both EC and ED approaches and satisfy equations~(\ref{eps-lan}) and (\ref{mu-lan}) in the limit of weakly absorbing medium.
Plugging equations~(\ref{eps-eff}) and (\ref{mu-eff}) into equation~(\ref{density}), we obtain the average EM energy density.
We emphasize that these calculations involving this particular set of parameters are an application of the analytic results presented in Secs.~\ref{mie} and \ref{time}, which are generally valid for other classes of dispersive, non-optically active metamaterials in time-harmonic fields.

\section{Numerical results}
\label{numerical}

\begin{figure}[htbp!]
\centering
\includegraphics[width=\columnwidth]{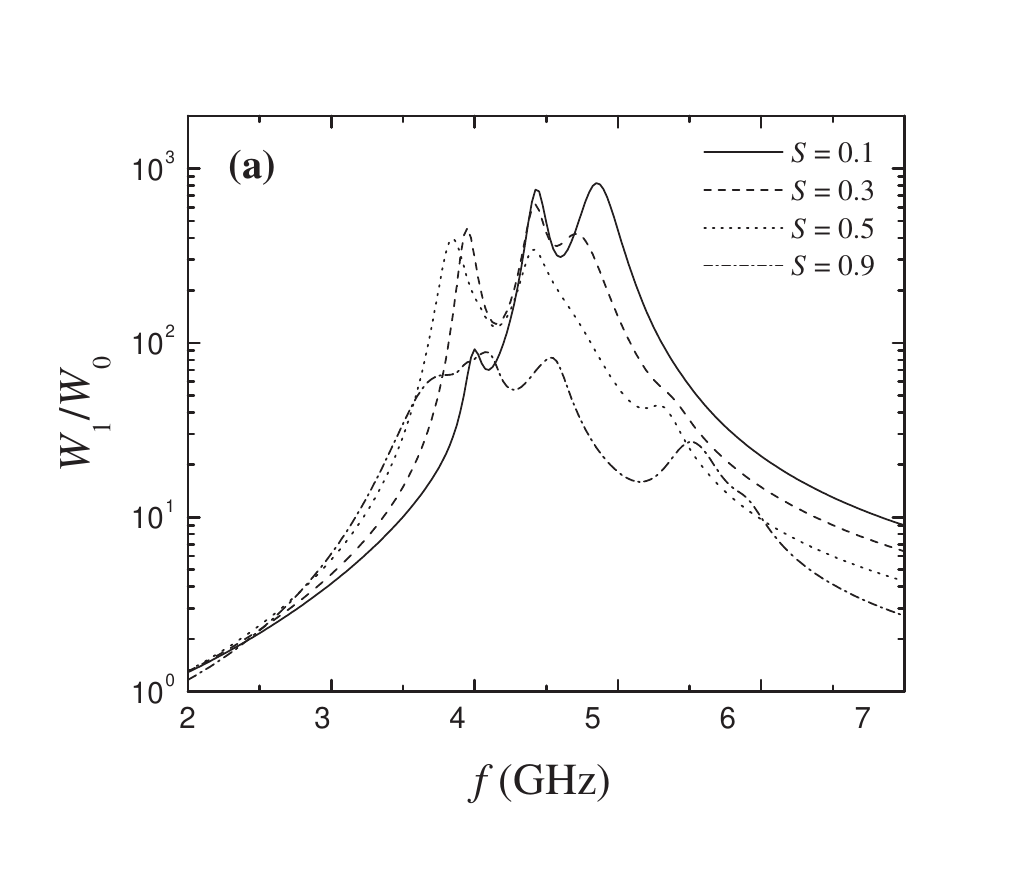}\vspace{-1cm}
\includegraphics[width=\columnwidth]{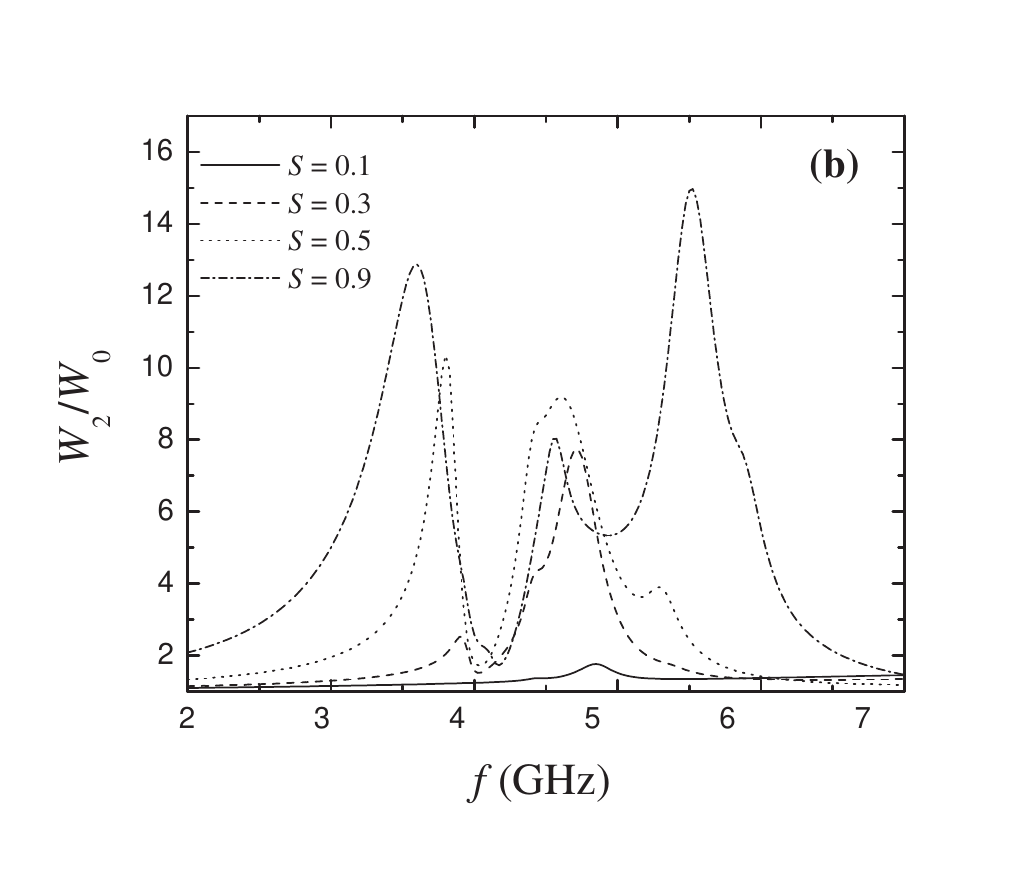}\vspace{-1cm}
\includegraphics[width=\columnwidth]{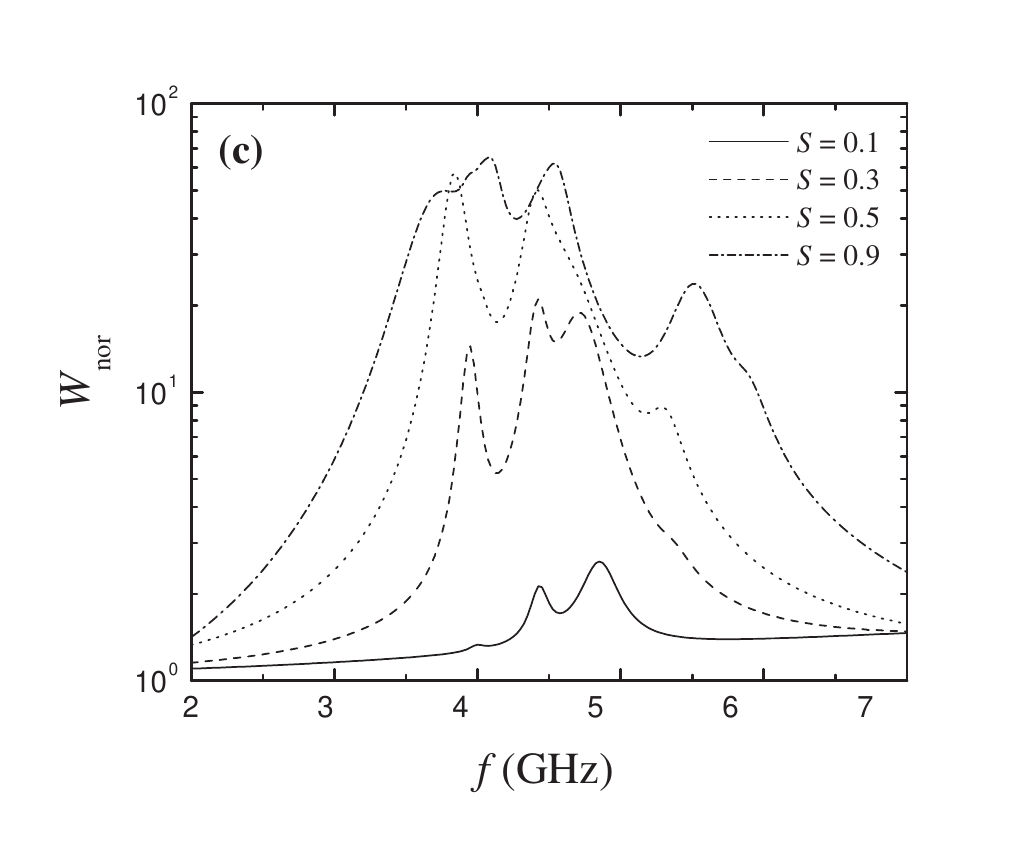}
\caption{Normalized EM energy inside a coated sphere with dispersive metamaterial core $[\epsilon_1(\omega),\mu_1(\omega)]$, given by equations~(\ref{eps1}) and (\ref{mu1}), and lossless dielectric shell $(\epsilon_2/\epsilon_0=1.6,\mu_2/\mu_0=1)$ as a function of the frequency $f=\omega/2\pi$, with thickness parameters $S=0.1,$ 0.3, 0.5, 0.9. {\bf (a)} EM energy $W_1(0,a)/W_0(0,a)$ inside the core $[\epsilon_1(\omega),\mu_1(\omega)]$. {\bf (b)} EM energy $W_2(a,b)/W_0(a,b)$ inside the shell $(\epsilon_2,\mu_2)$. {\bf (c)} EM energy $W_{\rm nor}=W_{1;2}/W_0$ inside the scatterer (core and shell). }
\label{fig1}
\end{figure}

To calculate the time-averaged EM energy as stated in equation~(\ref{W12}), it is suitable to deal with dimensionless quantities only.
From equations~(\ref{w0}), (\ref{W1}), (\ref{W2}) and (\ref{W12}), we obtain a normalization $W_{\rm nor}={W_{1;2}(a,b)}/{W_0(0,b)}$, which yields
\begin{eqnarray}
    W_{\rm nor}(S)=S^3\frac{W_1(0,a)}{W_0(0,a)}+(1-S^3)\frac{W_2(a,b)}{W_0(a,b)}\
    ,\label{W12-norm}
\end{eqnarray}
where $S=a/b$ is the thickness ratio of the coated sphere.
Note that $S^3$ and $(1-S^3)$ are the volume fraction of the spherical core and shell, respectively.
Here, all the numerical calculations have been performed by a computer code written for the free software {\it Scilab}~5.3.3.
According to Ruppin~\cite{Ruppin-negative}, we choose $f_{\rm p}=10$~GHz and $f_0=4$~GHz for the plasma and magnetic resonance frequencies, respectively.
The damping coefficients are assumed to be $\gamma = 0.03\omega_{\rm p}$ and $\Gamma=0.03\omega_0$, and the dimensionless parameter $F=0.56$ has been chosen.
As it is well-known, the effective parameters for metamaterials, equations~(\ref{eps1})--(\ref{mu-eff}), are valid provided the incident wavelength is much larger than the typical unit of the metamaterial (e.g. the split-ring resonator)~\cite{Soukoulisbook}.
As split-ring resonators are typically millimeter-sized, in all numerical calculations the frequency range has been limited to a few gigahertz to guarantee the validity of the expressions for the effective electromagnetic parameters.

With this set of parameters, the real parts of the electric permittivity and magnetic permeability are simultaneously negative for $f_0=4$~GHz to 6~GHz, so that a band of negative refraction shows up in this frequency range.
The extinction efficiency $Q_{\rm ext}$, which is the extinction cross-section in units of the geometric one, can be calculated as follows~\cite{Bohren}:
\begin{eqnarray}
    Q_{\rm ext}=\frac{2}{y^2}\sum_{n=1}^{\infty}(2n+1){\rm Re}\left(a_n+b_n\right)\ ,
\end{eqnarray}
where $y=kb$ is the size parameter of the outer sphere and $a_n$ and $b_n$ are the scattering coefficients given in equations~(\ref{an}) and (\ref{bn}).

In the following, we consider two typical cases as Gao and Huang have studied~\cite{Gao}: a coated sphere with dispersive core and dielectric shell, and vice versa.
The surrounding medium $(\epsilon_0,\mu_0)$ is assumed to be the vacuum and the radius $b=1$~cm is chosen for the outer sphere.
Note that, with this choice, we have $kb=\omega b({\epsilon_0\mu_0})^{1/2} \approx1$, for $f\approx5$~GHz.
For the dielectric and non-dispersive material, we consider the lossless parameters $\epsilon_q/\epsilon_0=1.6$ and $\mu_q/\mu_0=1$~\cite{Gao}.
As discussed in~\cite{Gao}, the curves of extinction efficiency become smoother when the dielectric shell or core are absorbing, leading to a decrease in the amplitude of $Q_{\rm ext}$ as the absorption increases for $S<0.5$ or $S>0.5$, respectively.
Here, we consider the lossless situation because the results when the core is dielectric and the shell is left-handed are quite similar to the one in which we have in the core vacuum or, more generally, the same lossless material as the surrounding medium.
Besides, the average EM energy in a non-dispersive medium can exactly be determined only for weakly absorbing or lossless media~\cite{Landau}.

\begin{figure}[htbp!]
\centering
\includegraphics[width=\columnwidth]{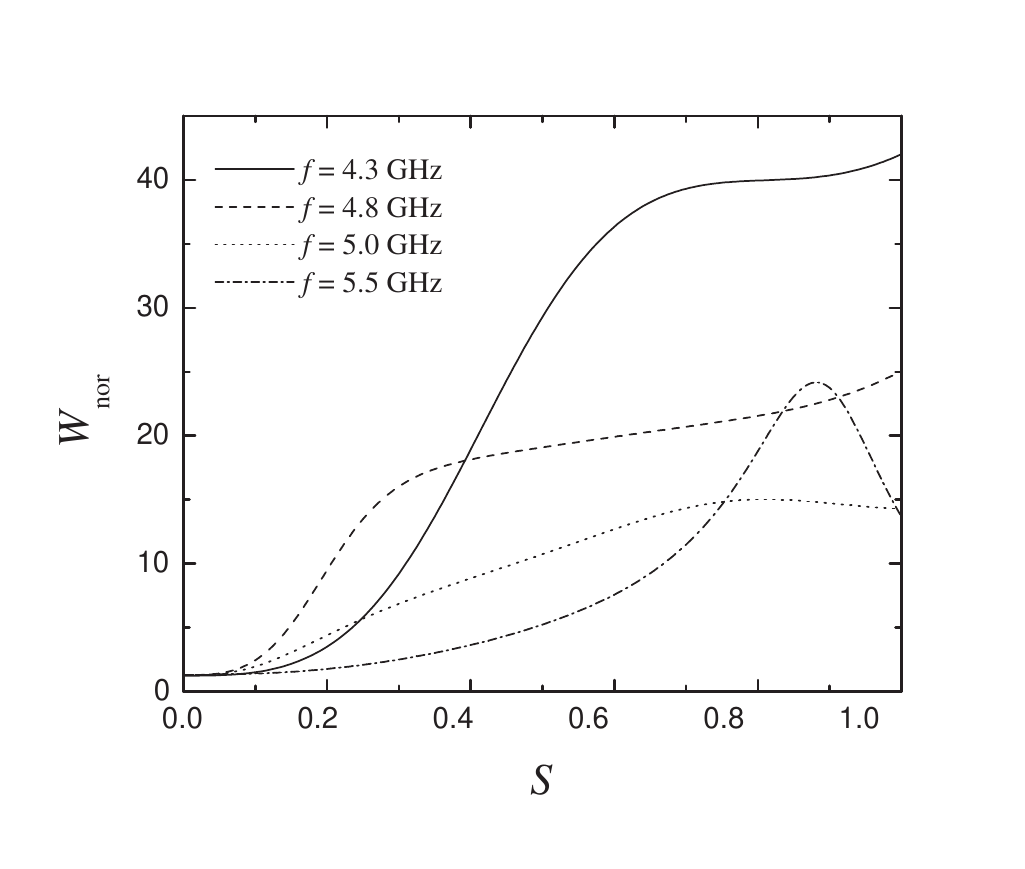}
\caption{Normalized EM energy $W_{\rm nor}=W_{1;2}/W_0$ inside a coated sphere with dispersive metamaterial core $[\epsilon_1(\omega),\mu_1(\omega)]$, given by equations~(\ref{eps1}) and (\ref{mu1}), and lossless dielectric shell $(\epsilon_2/\epsilon_0=1.6,\mu_2/\mu_0=1)$ as a function of the thickness ratio $S=a/b$ for some frequencies in the left-handed region. }
\label{fig2}
\end{figure}

Some curves of the average EM energy within the core [figure~\ref{fig1}(a)] and within the shell [figure~\ref{fig1}(b)], when the former is dispersive $[\epsilon_1=\epsilon_1(\omega),\mu_1=\mu_1(\omega)]$ and the latter is a lossless dielectric $(\epsilon_2/\epsilon_0=1.6,\mu_2/\mu_0=1)$, are presented as functions of the frequency and the thickness parameter.
In figure~\ref{fig1}(a), we observe in the left-handed region (4 to 6 GHz) a strong enhancement of the EM energy in the metamaterial core.
This is related to the presence of standing waves inside the particle.
Below and above this frequency range, the internal energy decreases monotonically.
This decrease is expected for right-handed frequencies [${\rm Re}(m_1)>0$], since there are no resonance peaks in the internal energy at the Rayleigh size parameters region ($ka<kb\leq1$) unless the scatterer exhibits high absolute values of permeability or permittivity~\cite{Tiago-sphere,Tiago-cylinder}, which is not the case for $f<3.5$~GHz and $f>6$~GHz.
Besides, once in this range $\epsilon_1'(\omega)<0$ or $\mu_1'(\omega)<0$, but not both, we have ${\rm Re}(m_1)\approx0$ for low absorption, and the EM waves cannot propagate inside the core, but just excite resonantly surface plasmon-polaritons, like the resonance peak between $3.5$~GHz and $4$~GHz~\cite{Ruppin-negative}.
These surface modes contributes to the enhancement of the EM energy within the dielectric shell, as can be observed in figure~\ref{fig1}(b).
With increasing the thickness ratio $S$, the energy within the dielectric shell increases with maximum values in the left- and right-handed regions, both associated with resonance surface modes~\cite{Ruppin-negative}.

\begin{figure}[htbp!]
\centering
\includegraphics[width=\columnwidth]{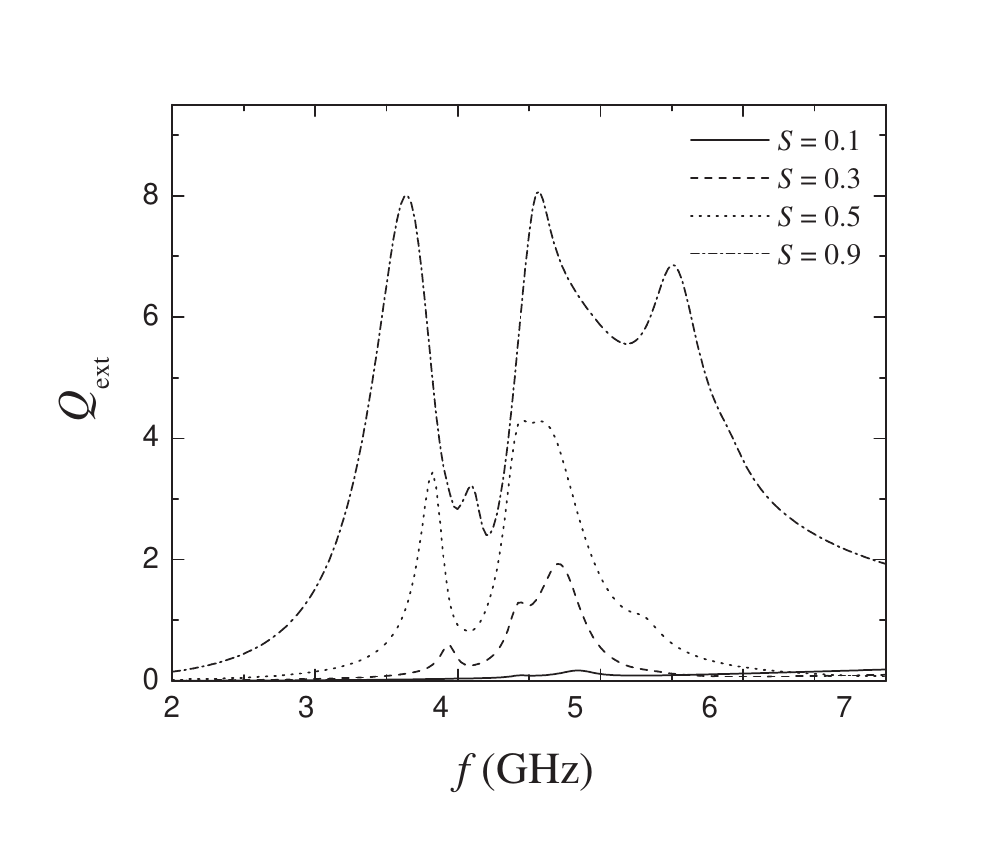}
\caption{Extinction efficiency $Q_{\rm ext}$ for a coated sphere with dispersive metamaterial core $[\epsilon_1(\omega),\mu_1(\omega)]$, given by equations~(\ref{eps1}) and (\ref{mu1}), and lossless dielectric shell $(\epsilon_2/\epsilon_0=1.6,\mu_2/\mu_0=1)$, with thickness parameters $S=0.1,$ 0.3, 0.5, 0.9. }
\label{fig3}
\end{figure}

\begin{figure}[htbp!]
\centering
\includegraphics[width=\columnwidth]{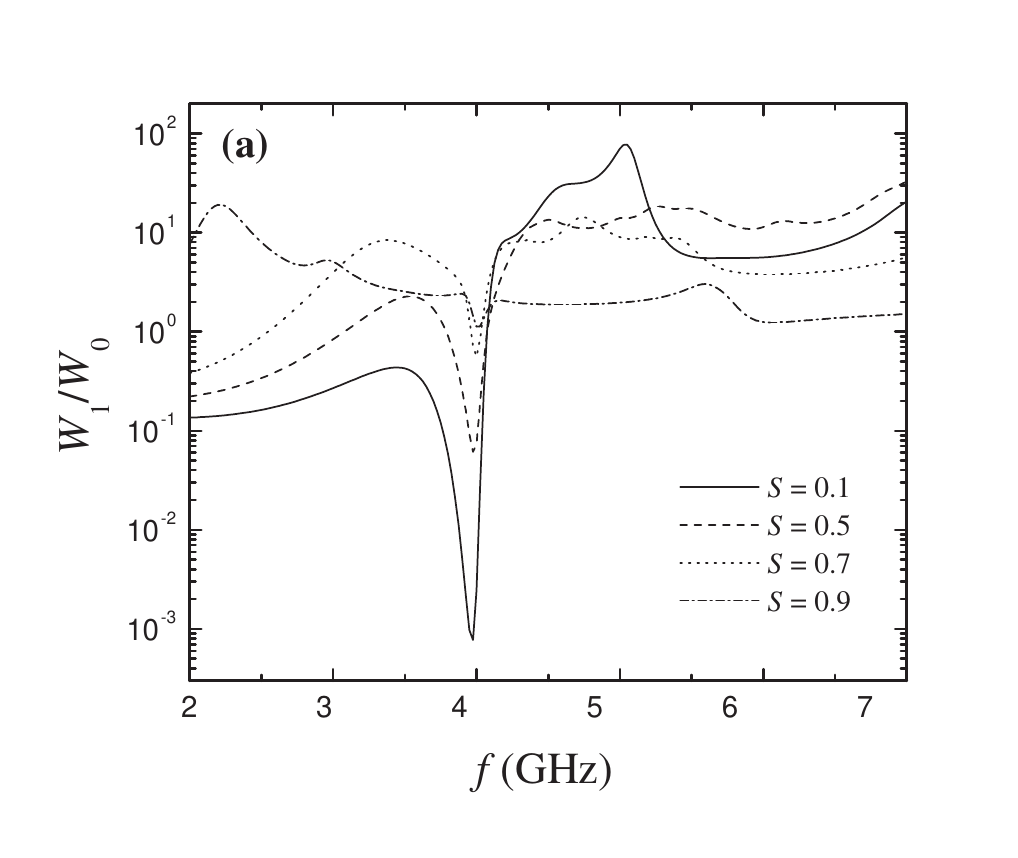}\vspace{-1cm}
\includegraphics[width=\columnwidth]{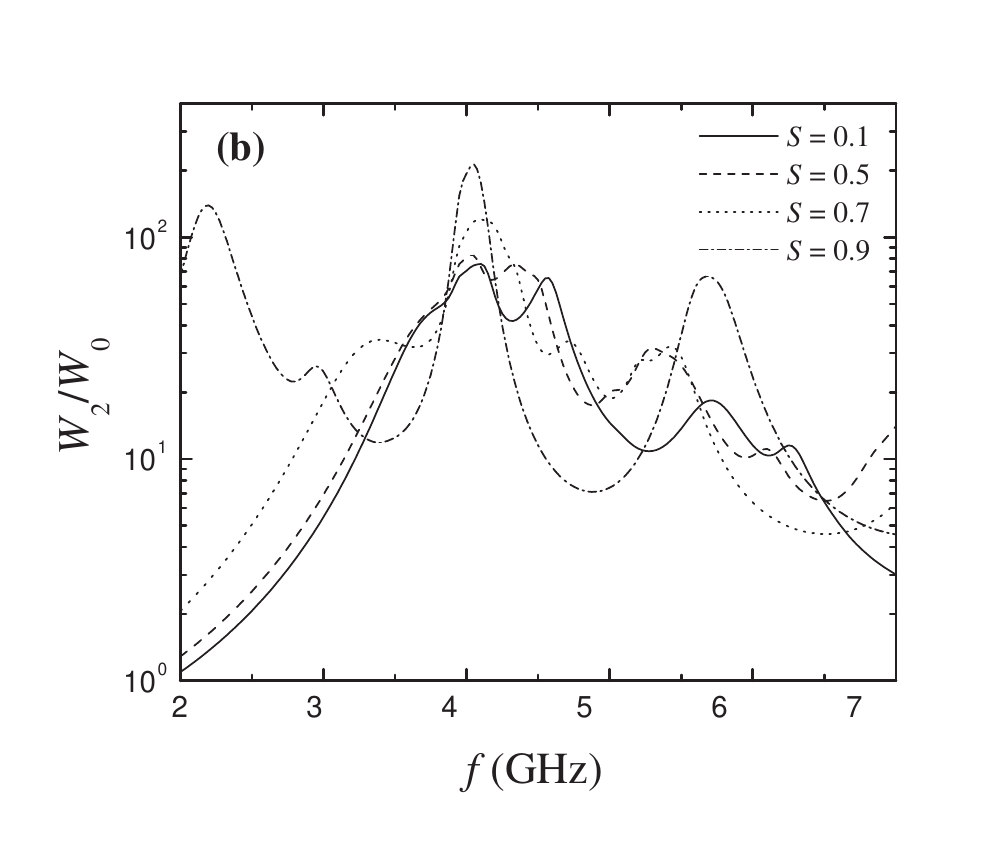}\vspace{-1cm}
\includegraphics[width=\columnwidth]{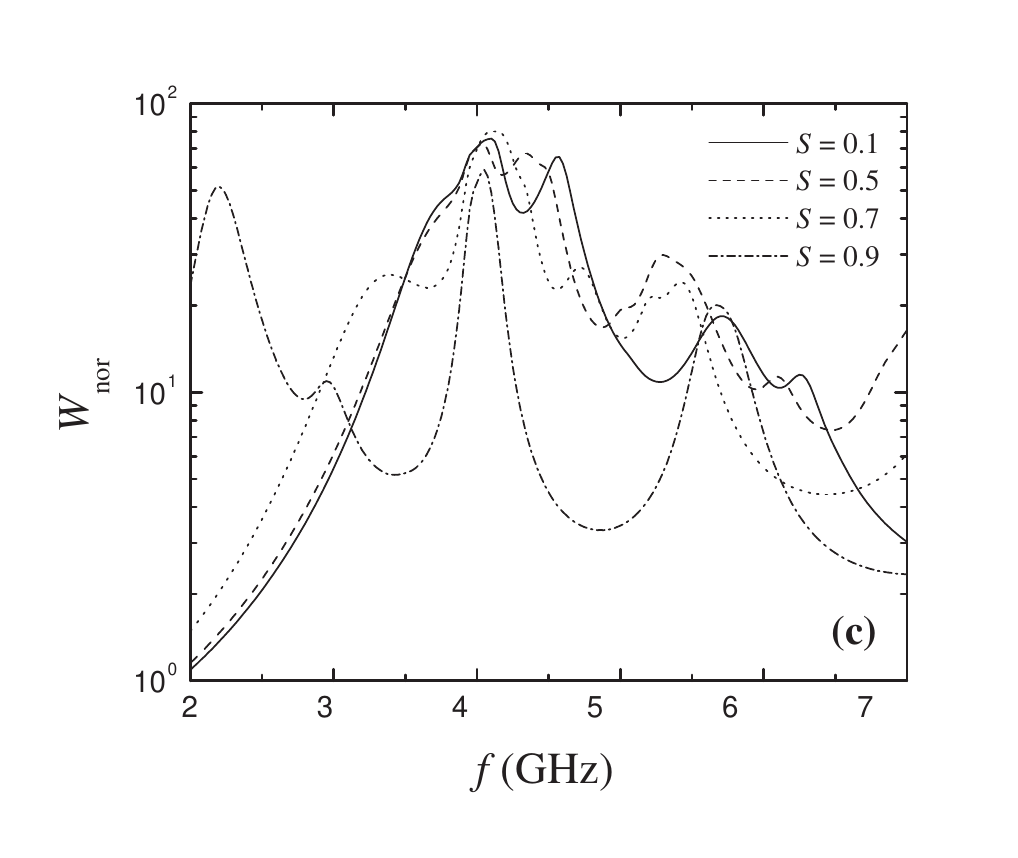}
\caption{Normalized EM energy inside a coated sphere with lossless dielectric core $(\epsilon_1/\epsilon_0=1.6,\mu_1/\mu_0=1)$ and dispersive metamaterial shell $[\epsilon_2(\omega),\mu_2(\omega)]$, given by equations~(\ref{eps1}) and (\ref{mu1}), as a function of the frequency $f=\omega/2\pi$, with thickness parameters $S=0.1,$ 0.5, 0.7, 0.9. {\bf (a)} EM energy $W_1(0,a)/W_0(0,a)$ inside the core $(\epsilon_1,\mu_1)$. {\bf (b)} EM energy $W_2(a,b)/W_0(a,b)$ inside the shell $[\epsilon_2(\omega),\mu_2(\omega)]$. {\bf (c)} EM energy $W_{\rm nor}=W_{1;2}/W_0$ inside the scatterer (core and shell).}
\label{fig4}
\end{figure}

Adding the EM energies $W_1(0,a)$ and $W_2(a,b)$ from figures~\ref{fig1}(a) and \ref{fig1}(b) by means of equation~(\ref{W12-norm}), we have obtained the profiles of the total EM energy within the coated sphere presented in figure~\ref{fig1}(c).
Note that the energy-enhancement factor $W_{\rm nor}$ is an increasing function at the left-handed region, becoming broader and with greater magnitude, as the thickness parameter $S$ is increased, which is the opposite of figure~\ref{fig1}(a).
This is explicitly shown in figure~\ref{fig2} for some frequencies in the left-handed range.
The same behavior can also be observed for the extinction efficiency, as we show in figure~\ref{fig3}.
These results for $b=1$~cm are quite different from those obtained in~\cite{Gao} for $b=10$~cm, where the amplitudes of the extinction efficiencies decrease with $S$.

In the extinction spectra, as discussed by Ruppin~\cite{Ruppin-negative}, the plasmon-like and the magnetic excitations reinforce each other in the low-frequency region below $\omega_0$.
In this region, the real part of the permittivity is negative, whereas the permeability has positive real part [$\epsilon_1'(\omega)<0,\mu_1'(\omega)>0]$.
Since these two mechanisms can be considered as roughly independent, the peaks below $\omega_0$ are provided by the contributions of the magnetic bulk polaritons and surface plasmon-polaritons, where these latter are due to resonances in the Aden-Kerker coefficient $a_n$.
However, just above $\omega_0$, both the electric permittivity and magnetic permeability have negative real parts, and the surface plasmon-polaritons and magnetic surface polaritons, which are due to resonances in the multipole moments $a_n$ and $b_n$, respectively, suppress each other.
Physically, this happens because, except for small intrinsic absorption, the metamaterial becomes transparent in this left-handed region and, thereby, no surface modes can exist~\cite{Ruppin-negative}.

Now we consider the reciprocal situation: the coated sphere is made of a lossless dielectric core ($\epsilon_1/\epsilon_0=1.6,\mu_1/\mu_0=1$) and a wires-SRR metamaterial shell $[\epsilon_2=\epsilon_2(\omega),\mu_2=\mu_2(\omega)]$, given by equations~(\ref{eps1}) and (\ref{mu1}).
Once again, we assume $b=1$~cm and consider the same parameters we have used above for the dispersive metamaterial, leading a left-handed range of frequencies 4 to 6~GHz.
Curves of the time-averaged EM energy within the spherical core and shell as functions of the incident frequency, for some thickness ratios, are presented in figures~\ref{fig4}(a) and \ref{fig4}(b), respectively.

\begin{figure}[htbp!]
\centering
\includegraphics[width=\columnwidth]{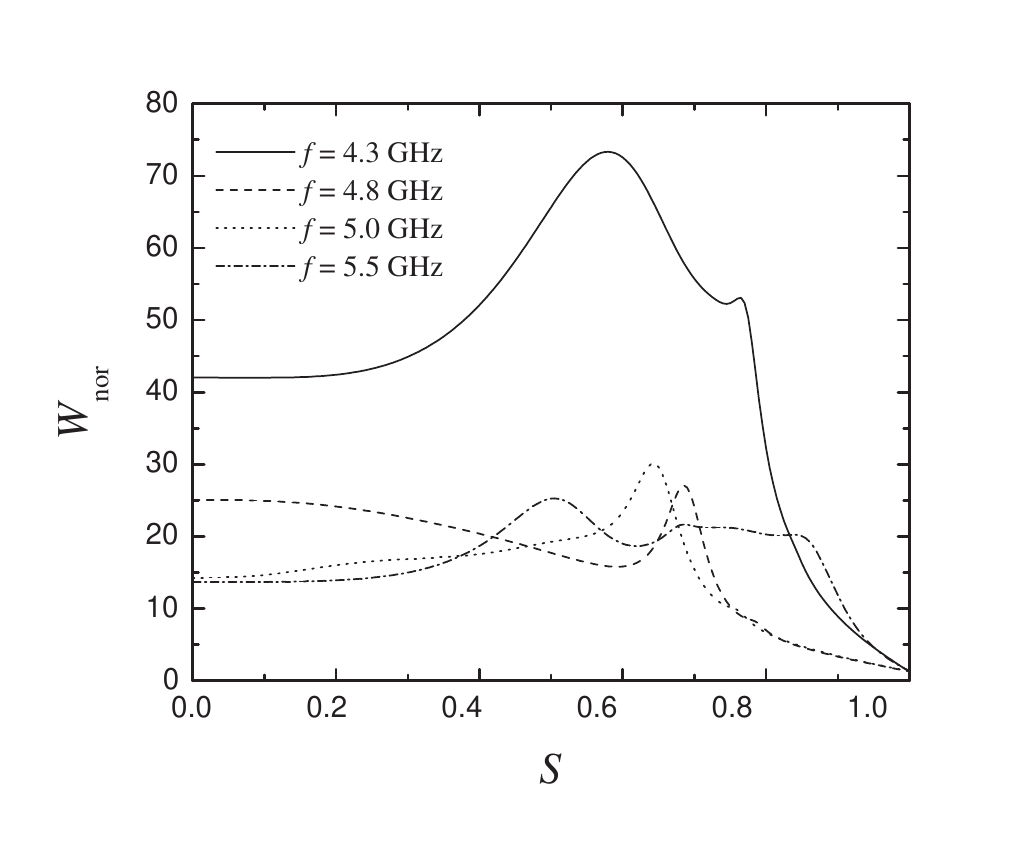}
\caption{Normalized EM energy $W_{\rm nor}=W_{1;2}/W_0$ inside a coated sphere with lossless dielectric core $(\epsilon_1/\epsilon_0=1.6,\mu_1/\mu_0=1)$ and dispersive metamaterial shell $[\epsilon_1(\omega),\mu_1(\omega)]$, given by equations~(\ref{eps1}) and (\ref{mu1}), as a function of the thickness ratio $S=a/b$ for some frequencies in the left-handed region.}
\label{fig5}
\end{figure}

\begin{figure}[htbp!]
\centering
\includegraphics[width=\columnwidth]{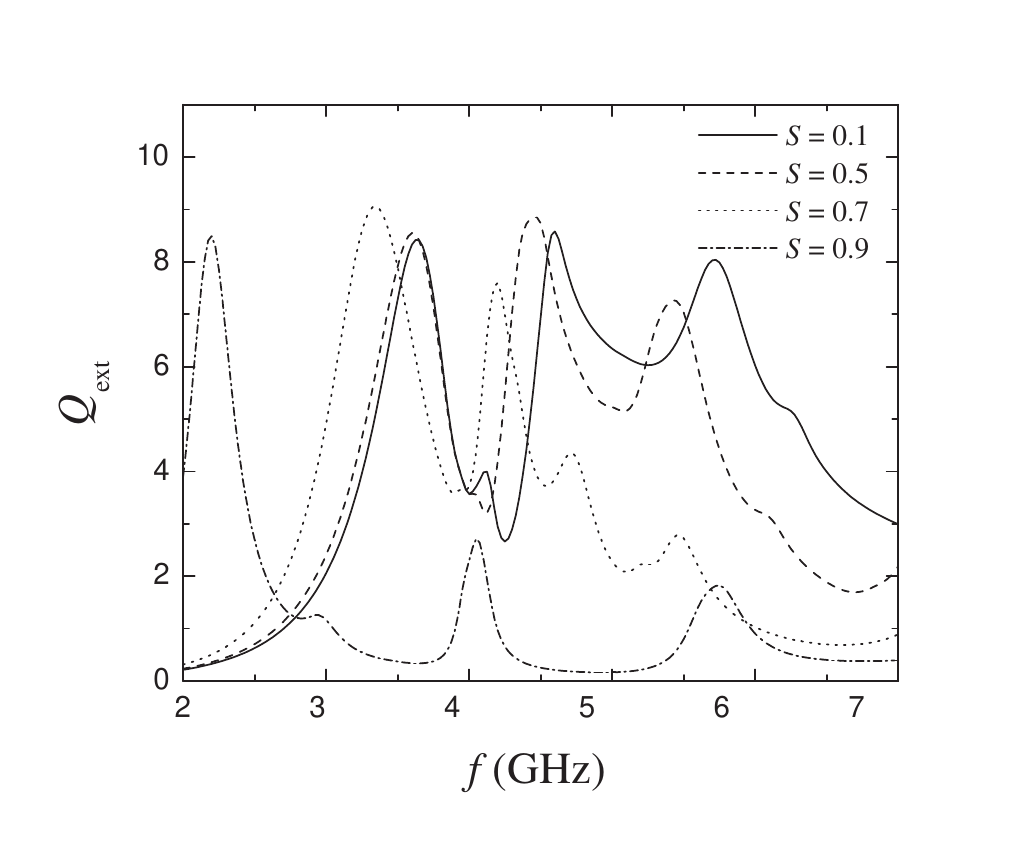}
\caption{Extinction efficiency $Q_{\rm ext}$ for a coated sphere with lossless dielectric core $(\epsilon_1/\epsilon_0=1.6,\mu_1/\mu_0=1)$ and dispersive metamaterial shell $[\epsilon_2(\omega),\mu_2(\omega)]$, given by equations~(\ref{eps1}) and (\ref{mu1}), with thickness parameters $S=0.1,$ 0.5, 0.7, 0.9. }
\label{fig6}
\end{figure}

In figure~\ref{fig4}(b), we note that the resonance peaks inside the metamaterial shell are shifted to low-frequencies with increasing the amount of dielectric material within the coated sphere.
Specially, the EM energy within the dielectric core is strongly suppressed in the low-frequency range $\omega<\omega_0$, with minimum in $f_0=4$~GHz [figure~\ref{fig4}(a)].
This behavior is due to the excitation of surface plasmon-polaritons in the metamaterial shell at this frequency range, which provides highest values of EM field intensity at the surface of the sphere and they decay evanescently towards its center.
Thereby, for ``skin depth'' smaller than the layer thickness $(b-a)$, the EM field barely reaches the dielectric core, and this is clearly seen in figure~\ref{fig4}.
It is important to point out that this result can only be obtained by means of a direct computation of the stored EM fields in the core and shell separately, emphasizing the importance of the present analysis.
The total EM energy inside the coated sphere is shown in figures~\ref{fig4}(c) and \ref{fig5}, where we can observe a decrease of the EM energy in the left-handed region with the thickness parameter $S$.
The same behavior is also found in the extinction spectra, as we show in figure~\ref{fig6}.

Finally, let us briefly discuss some possible applications for our findings.
Coated spheres are frequently a key element in plasmonic devices and metamaterials that support surface plasmon resonances, which have a wide range of applications, ranging from biomedical imaging, surface-enhanced spectroscopies, chemical and biological sensing to the metamaterial analog of electromagnetically induced transparency~\cite{Fanoreview,Fanoshells}.
These applications essentially rely on two distinctive EM properties of coated spheres: the Fano resonances and the great enhancement of the EM field in the vicinities of a plasmon resonance.
Recently, it has been shown that metamaterial spheres with negative refraction are more likely to exhibit Fano resonances in Lorenz-Mie scattering than conventional materials, so that metamaterial can be considered to be ``predisposed'' to possessing Fano resonances~\cite{Fanoreview}.
This fact makes our results potentially interesting for applications specially because the energy density in coated metamaterials, which we explicitly calculate here, is an observable sensitive to the interference between electromagnetic resonances, the ultimate cause of the Fano effect.
Indeed, the energy density is not just proportional to the square of the Aden-Kerker coefficients (intensities), as other light scattering observables such as the extinction and scattering efficiencies, making them insensitive to interference effects~\cite{Fanoreview}.
Rather, it depends on cross-terms involving different Aden-Kerker coefficients, as we have unveiled and is evident from the inspection of equations~(\ref{w2-e}) and~(\ref{w2-h}).
As a result, the stored EM energy could be explored not only to probe Fano resonances in applications, which in addition will be more likely to occur due to negative refraction in metamaterials, but also to identify regions of field enhancement.
We have numerically obtained that the absorption cross-section (in units of the geometric cross-section) is related to the stored electric and magnetic energies by means of the equation
\begin{widetext}
    \begin{eqnarray}
        Q_{\rm abs} &=&
        \frac{8}{3}y\Bigg\{\left[\frac{\epsilon_1''}{\epsilon_1^{(\rm eff)}}\frac{W_{1E}(0,a)}{W_0(0,a)}
        + \frac{\mu_1''}{\mu_1^{(\rm
        eff)}}\frac{W_{1H}(0,a)}{W_0(0,a)}\right]S^3 + \left[\frac{\epsilon_2''}{\epsilon_2^{(\rm eff)}}\frac{W_{2E}(a,b)}{W_0(a,b)}
        + \frac{\mu_2''}{\mu_2^{(\rm
        eff)}}\frac{W_{2H}(a,b)}{W_0(a,b)}\right](1-S^3)\Bigg\}\ , \label{link}
    \end{eqnarray}
\end{widetext}
which extends the result already obtained by Ruppin~
\cite{Ruppin-energy} and can be straightforwardly generalized to multilayered spheres.
Equation (\ref{link}) puts in evidence the connection that exists between the calculated observable, the stored energy inside the coated sphere, with a directly measurable quantity, the absorption efficiency.
It is worth recalling that the EM field stored in coated spheres made of metamaterials, and hence Fano resonances, sensitively depends on the specific frequency dependence of the effective parameters $\epsilon_q (\omega)$ and $\mu_q (\omega)$~\cite{Fanoreview}.
This fact just reinforces the importance of considering realistic effective parameters for metamaterials [equations~(\ref{eps1})--(\ref{mu-eff})].

\section{Conclusions}
\label{conclusions}

We have analytically calculated an exact expression for the time-averaged EM energy within a coated sphere, irradiated by a plane wave, using the explicit expressions for the EM fields inside the core and the shell separately.
Although one can infer some properties of the internal energy and power loss in metamaterial coated spheres from the analysis of the extinction and scattering efficiencies, a direct computation of the stored EM fields in the core and shell separately allows a complete understanding of the behavior of EM energy density in such systems.
Here, we have shown some profiles of the internal energy for two situations involving a metamaterial coated sphere: the dielectric shell and dispersive metamaterial core, and vice-versa.
For the metamaterial, we have used realistic dispersion relations for dispersive and lossy split-ring resonators.
To link the calculated quantities to measurable ones, we have numerically obtained an explicit relation between the stored electric and magnetic energies and the corresponding absorption efficiency for coated spheres.
We have shown that the stored EM energy explicitly depends on cross-terms involving different Aden-Kerker coefficients.
This finding unveils the fact that the stored energy is an observable sensitive to field interferences responsible for the Fano effect.
Since the Fano effect is more likely to occur in the presence of negative refraction~\cite{Fanoreview}, we suggest that the stored energy in coated spheres could probe Fano resonances in metamaterials.

\section*{Acknowledgments}

The authors acknowledge the Brazilian agencies for support.
TJA holds grants from Fundação de Amparo à Pesquisa do Estado de São Paulo (FAPESP) (2008/02069-0 and 2010/10052-0), ASM from Conselho Nacional de Desenvolvimento Científico e Tecnológico (CNPq) (305738/2010-0 and 476722/2010-1), and FAP from Fundação de Amparo à Pesquisa do Estado do Rio de Janeiro (FAPERJ) (E-26/111.463/2011).


\begin{thebibliography}{99}



\bibitem{Bohren}
C. F. Bohren and D. R. Huffman,
{\it Absorption and Scattering of Light by Small Particles}
(Wiley, 1983).

\bibitem{Hulst}
H. C. van de Hulst, ``Light Scattering by Small Particles''
(Dover, 1981).

\bibitem{Tribelsky}
M. I. Tribelsky and B. S. Luk'yanchuk,
``Anomalous light Scattering by small particles,''
Phys. Rev. Lett. \textbf{97}, 263902 (2006).


\bibitem{Bashevoy}
M. Bashevoy, V. Fedotov, and N. Zheludev,
``Optical whirlpool on an absorbing metallic nanoparticle,''
Opt. Express \textbf{13}, 8372--8379 (2005).


\bibitem{Ruan}
Z. Ruan and S. Fan,
``Superscattering of light from subwavelength nanostructures,''
Phys. Rev. Lett. \textbf{105}, 013901 (2010).


\bibitem{Maier}
S. A. Maier,
{\it Plasmonics: Fundamentals and Applications}
(Springer, 2007).

\bibitem{Aden}
A. L. Aden and M. Kerker,
``Scattering of electromagnetic waves from two concentric spheres,''
J. Appl. Phys. {\bf 22}, 1242--1246 (1951).



\bibitem{Miroshnichenko2010}
A. E. Miroshnichenko,
``Off-resonance field enhancement by spherical nanoshells,''
Phys. Rev. A {\bf 81}, 053818 (2010).


\bibitem{Spaser}
M. A. Noginov, G. Zhu, A. M. Belgrave, R. Bakker, V. M. Shalaev, E. E. Narimanov, S. Stout, E. Herz, T. Suteewong, and U. Wiesner,
``Demonstration of a spaser-based nanolaser,''
Nature {\bf 460}, 1110--1112 (2009).



\bibitem{Smith00}
D. R. Smith, W. J. Padilla, D. C. Vier, S. C. Nemat-Nasser, and S. Schultz,
``Composite medium with simultaneously negative permeability and permittivity,''
{Phys. Rev. Lett.} {\bf 84}, 4184--4187 (2000).

\bibitem{Smith2004}
D. R. Smith, J. B. Pendry, and M. C. K. Wiltshire,
``Metamaterials and negative refractive index,''
Science \textbf{305}, 788--792 (2004).

\bibitem{Enkrich2005}
C. Enkrich, M. Wegener, S. Linden, S. Burger, L. Zschiedrich, F. Schmidt, J. F. Zhou, Th. Koschny, and C. M. Soukoulis,
``Magnetic metamaterials at telecommunication and visible frequencies,''
Phys. Rev. Lett. \textbf{95}, 203901 (2005).

\bibitem{Cai2007}
W. Cai, U. K. Chettiar, H. K. Yuan, V. C. de Silva, A. V. Kildishev, V. P. Drachev, and V. M. Shalaev,
``Metamagnetics with rainbow colors,''
Opt. Express \textbf{15}, 3333--3341 (2007).

\bibitem{Pendry2006}
J. B. Pendry, D. Shurig, and D. R. Smith,
``Controlling electromagnetic fields,''
Science \textbf{312}, 1780--1782 (2006).

\bibitem{Leonhardt2006}
U. Leonhardt,
``Optical conformal mapping,''
Science \textbf{312}, 1777--1780 (2006).


\bibitem{Ruppin-negative}
R. Ruppin,
``Extinction properties of a sphere with negative permittivity and permeability,''
{Solid State Comm.} {\bf 116}, 411--415 (2000).

\bibitem{Gao}
L. Gao and Y. Huang,
``Extinction properties of a coated sphere containing a left-handed material,''
Opt. Comm. {\bf 239}, 25--31 (2004).


\bibitem{Liu2004}
Z. Liu, Z. Lin, and S. T. Chui,
``Electromagnetic scattering by spherical negative-refractive-index particles: Low-frequency resonance and localization parameters,''
Phys. Rev. E {\bf 69}, 016619 (2004).

\bibitem{Tretyakov2005}
S. A. Tretyakov, S. I. Maslovski, A. A. Sochava, and C. R. Simovski, ``The influence of complex material coverings on the quality factor of simple radiating systems,''
IEEE Trans. Antennas Propagat. {\bf 53}, 965--970 (2005).

\bibitem{Tretyakov}
S. A. Tretyakov,
``Electromagnetic field energy density in artificial microwave materials with strong dispersion and loss,''
Phys. Lett. A {\bf 343}, 231--237 (2005).

\bibitem{Boardman}
A. D. Boardman and K. Marinov,
``Electromagnetic energy in a dispersive metamaterial,''
Phys. Rev. B {\bf 73}, 165110 (2006).

\bibitem{Luan1}
P. G. Luan,
``Power loss and electromagnetic energy density in a dispersive metamaterial medium,''
Phys. Rev. E {\bf 80}, 046601 (2009).

\bibitem{Luan2}
P. G. Luan, Y. T. Wang, S. Zhang, and X. Zhang,
``Electromagnetic energy density in a single-resonance chiral metamaterial,''
Opt. Lett. {\bf 36}, 675--677 (2011).

\bibitem{Bott}
A. Bott and W. Zdunkowski,
``Electromagnetic energy within dielectric spheres,''
J. Opt. Soc. Am. A {\bf 4}, 1361--1365 (1987).

\bibitem{Ruppin-sphere}
{R. Ruppin},
``Electromagnetic energy in dispersive spheres,''
{J. Opt. Soc. Am. A} {\bf 15}, 524--527 (1998).

\bibitem{Tiago-sphere}
T. J. Arruda and A. S. Martinez,
``Electromagnetic energy within a magnetic sphere,''
J. Opt. Soc. Am. A {\bf 27}, 992--1001 (2010).

\bibitem{Ruppin-energy}
R. Ruppin,
``Electric and magnetic energies within dispersive metamaterial spheres,''
J. Opt. {\bf 13}, 095101 (2011).


\bibitem{Kerker:1983}
M.~Kerker, D.~S. Wang, and C.~L. Giles,
``Electromagnetic scattering by magnetic spheres,''
J. Opt. Soc. Am. {\bf 73}, 765--767 (1983).

\bibitem{Alexandre-new}
F. A. Pinheiro, A. S. Martinez, and L. C. Sampaio,
``New effects in light scattering in disordered media and coherent backscattering cone: system of magnetic particles,''
Phys. Rev. Lett. {\bf 84}, 1435--1438 (2000).

\bibitem{Alexandre-vanish}
F. A. Pinheiro, A. S. Martinez, and L. C. Sampaio,
``Vanishing of energy transport velocity and diffusion constant of electromagnetic waves in disordered magnetic media,''
Phys. Rev. Lett. {\bf 85}, 5563--5566 (2000).

\bibitem{Landau}
L. D. Landau and E. M. Lifshitz,
{\it Electrodynamics of Continuous Media}
(Pergamon, 1984).

\bibitem{Watson}
G. N. Watson,
{\it A Treatise on the Theory of Bessel Functions}
(Cambridge Univ. Press, 1958).

\bibitem{Tiago-cylinder}
T. J. Arruda and A. S. Martinez,
``Electromagnetic energy within a magnetic infinite cylinder and scattering properties for oblique incidence,''
J. Opt. Soc. Am. A {\bf 27}, 1679--1687 (2010).


\bibitem{Kaiser}
T. Kaiser, S. Lange, and G. Schweiger,
``Structural resonances in a coated sphere: investigation of the volume-averaged source function and resonance positions,"
Appl. Opt. {\bf 33}, 7789--7797 (1994).

\bibitem{Soukoulisbook}
P. Markos and C. M. Soukoulis,
{\it Wave Propagation: From Electrons to Photonic Crystals and
Left-Handed Materials} (Princeton Univ. Press, 2008).

\bibitem{Loudon}
``The propagation of electromagnetic energy through an absorbing
dielectric,'' J. Phys. A: Gen. Phys. {\bf 3}, 233--245 (1970).

\bibitem{Ruppin-dispersive}
R. Ruppin, ``Electromagnetic energy density in a dispersive and
absorptive material'' Phys. Lett. A. {\bf 299}, 309--312 (2002).

\bibitem{Fanoreview}
B. Luk'yanchuk, N. I. Zheludev, S. A. Maier, N. J. Halas, P. Nordlander, H. Giessen, and C. T. Chong
``The Fano resonance in plasmonic nanostructures and
metamaterials,'' Nature Materials {\bf 9}, 707--715 (2010).

\bibitem{Fanoshells}
S. Mukherjee, H. Sobhani, J. B. Lassiter, R. Bardhan, P. Nordlander, and N. J. Halas,
``Fanoshells: nanoparticles with built-in Fano resonances,'' Nano
Lett. {\bf 10}, 2694--2701 (2010).

\end{thebibliography}
\end{document}